\documentclass[english,a4paper]{article}

\usepackage{graphicx}
\usepackage{float}
\usepackage{amssymb,amsfonts,amsmath}
\usepackage{mathrsfs}
\usepackage{color}
\usepackage{colordvi}
\usepackage{graphicx}
\usepackage[sans]{dsfont}
\usepackage{esvect}
\usepackage{enumerate}

\newtheorem{proposition}{Proposition}

\definecolor{MyGreen}{rgb}{0.1,0.6,0}

\newcommand{\T}{{\rm T}_{\rm diff}}
\renewcommand{\d}{{}_{\Delta}}
\newcommand{\dfn}[1]{\textit{#1}}

\newcommand{\R}{\mathbb R}
\newcommand{\Z}{\mathbb Z}

\newcommand{\mn}{\mathnormal}
\newcommand{\mrm}{\mathrm}
\renewcommand{\H}{\mathscr H}
\newcommand{\e}{\mathrm e}
\newcommand{\SU}{\mathrm{SU}}
\newcommand{\SO}{\mathrm{SO}}

\newcommand{\f}{^{*}}

\renewcommand{\f}{^*}
\newcommand{\thf}{\theta^*}
\newcommand{\psif}{\psi^*}
\newcommand{\phif}{\phi^*}

\newcommand{\tf}{t\f}

\newcommand{\til}{\widetilde}

\newcommand{\bl}{{}_{\bullet}}
\renewcommand{\i} {\mathbf{i}}
\renewcommand{\j} {\mathbf{j}}
\renewcommand{\k} {\mathbf{k}}
\newcommand{\id}{\mathds{1}}

\newcommand{\p}{\mathrm p}

\newcommand{\pf}{\p^0f^0}

\newcommand{\B}{\mrm B}
\newcommand{\rot}{\mathcal R}
\newcommand{\sgn}{\mathrm{sgn}}
\newcommand{\rhoo}{\rho_0}

\newcommand{\End}{\mathrm{End}}
\renewcommand{\Psi}{\boldsymbol\psi}
\renewcommand{\Theta}{\boldsymbol\theta}
\renewcommand{\Phi}{\boldsymbol\phi}
\renewcommand{\Lambda}{\boldsymbol\lambda}
\newcommand{\U}{\boldsymbol U}

\begin{document}
\title{Time-optimal control of SU(2) quantum operations}

\author{A. Garon, S. J. Glaser\footnote{Department of Chemistry, Technische Universit\"at
M\"unchen, Lichtenbergstrasse 4, D-85747 Garching, Germany}, D. Sugny\footnote{Laboratoire Interdisciplinaire Carnot de
Bourgogne (ICB), UMR 6303 CNRS-Universit\'e de Bourgogne, 9 Av. A.
Savary, BP 47 870, F-21078 DIJON Cedex, FRANCE, dominique.sugny@u-bourgogne.fr}}

\maketitle
\begin{abstract}
We propose an analysis of the time-optimal control of SU(2) quantum operations. By using the Pontryagin Maximum Principle, we show how to determine the optimal trajectory reaching a given target state. Explicit analytical solutions are given for two specific examples. We discuss the role of the detuning in the construction of the optimal synthesis.
\end{abstract}

\section{Introduction}
Manipulating a quantum system by an external field to achieve a given task remains a primary goal of various areas \cite{reviewQC} extending from atomic and molecular physics \cite{rice,shapiro,tannorbook,viellard}, Nuclear Magnetic Resonance \cite{spin}, quantum computing \cite{nielsen} to solid state physics. One of the most general and versatile procedures to tackle such control problems is optimal control theory \cite{pont,bonnardbook}. This technique designs a field able to bring the quantum system to the target state while maximizing or minimizing a given cost functional.

In our setting, an optimal control problem can be treated by two different types of approaches, geometric \cite{lapertglaser,boscain,alessandro,3level,khaneja,khaneja1,BS,BCS} and numerical methods \cite{grapetheory,skinner1,reich,gross,kosloff} for quantum systems of low and high dimension, respectively.
In addition to the problem of steering the density operator of the system to the target state, the creation of desired unitary operators plays a key role in both spectroscopy and quantum information processing.
Various approaches to construct a desired unitary transformation has been proposed.
One of them is based on the combination of simple pulses according to general symmetry principles \cite{spin,Tycko,levitt0,Luy} but the control fields constructed with this method generally present the drawback of having long durations.
Time-optimal solutions can be determined using numerical optimal control methods \cite{grapetheory,Palao,schulteSporl,JanichSchulte,JanichMcLean,kobzar,salomon} as well as using geometrical principles \cite{KhanejaHeitmann,khaneja1}. Note that related problems can also be solved by the use of Lie group techniques \cite{boscainchitour}.

This work will use geometric principles to address one of the basic and fundamental question in quantum computation: the optimal control of a single-qubit gate.
Up to now, this well-known control problem has been the subject of a series of works both in mathematics \cite{kirillova,chitour} and in physics \cite{fonseca,wenin,wu,boozer}, to cite a few.
Here, we propose to revisit this question by giving an explicit coordinate-parametrization of the optimal trajectories.
This leads to a complete, analytical and straightforward resolution of the control problem.
We will also discuss the influence of a detuning term on the optimal pulse sequence.

Here we consider a spin-$\frac12$ particle interacting with a constant magnetic field $\vec \B_0$ along the $z$-direction.
The control of the spin is performed through the variation of an external transverse magnetic field $\vec \B_1(t)$.
We place ourselves in the framework of the rotating wave approximation, for which $|\vec\B_1|<<|\vec\B_0|$ and the frequency of the transverse magnetic field is close to the Larmor frequency of the spin \cite{spin,Rabi1,Rabi2}.

To simplify the discussion, we consider in this paper only gates on SU(2) and not on U(2). We recall that the elements of SU(2) are the matrices of U(2) with determinant equal to one, i.e. the elements for which a global phase factor has been removed. The formalism introduced below can be straightforwardly extended to the unitary group U(2). For a given target state belonging to SU(2), we determine the control fields which minimize the total time of the process by applying the Pontryagin Maximum Principle.
Note that a similar control problem has been recently treated in Ref. \cite{boozer} in which a different parameterization of the optimal trajectories has been used. Our work complements this first study by addressing related questions such as the optimization on SO(3) as well as the computation of trajectories when a detuning term is taken into account.

The remainder of the paper is organized as follows.
The model of the system is presented in Sec. \ref{sec2} and the different choices of coordinates to parametrize SU(2) are discussed.
In Sec. \ref{sec3}, we show how to apply the Pontryagin Maximum Principle to this quantum system.
Sections \ref{sec4} and \ref{sec5} are devoted to the computation of the optimal trajectories with and without detuning.
Special attention is paid to two specific examples, the rotations about the $z$- axis and the rotations about axes in the ($x,y$)- plane. Explicit optimal solutions are given for these two quantum operations.
Conclusion and prospective views are given in Sec. \ref{sec6}.
Finally, some technical computations are reported in Appendix \ref{app}.

\section{The model system}%
\label{sec2}

This section is dedicated to presenting the general problem studied throughout the paper.
After introducing the system under concern, we will describe the problem mathematically and translate it in different coordinate systems in order to get several points of view, which will be used in the analysis of the optimal control problem.

\subsection{Spin Systems}
We consider a one spin-$\frac12$ closed system on which a constant magnetic field $\vec\B_0$ aligned in the $z$-direction is applied.
In addition, the system can be acted upon by a controlled transverse magnetic field $\vec\B_1(t)$ of bounded strength \cite{spin}.
As mentioned in the introduction, we assume that $|\vec \B_1|$ is much smaller than $|\vec \B_0|$ and that this field oscillates at a frequency close to the Larmor frequency of the spin.
Under this hypothesis, we can place ourselves in a given rotating frame and use the rotating wave approximation in order to simplify the description of the problem.
In this framework, the time-dependant quantum Hamiltonian of the system takes the form
$$
\mathrm H = \omega_x(t)\frac{\sigma_x}{2}+\omega_y(t)\frac{\sigma_y}{2}+\omega\frac{\sigma_z}{2},
$$
where $\omega_x$, $\omega_y$ are the control fields, which satisfy $\omega_x^2+\omega_y^2\leq\omega_{max}^2$.
The constant $\omega$ is the \dfn{detuning} term and corresponds (up to a constant factor) to the frequency difference between the frequency of the control field $\vec \B_1(t)$ and the Larmor frequency.
In particular, $\omega$ is zero in the resonant case where the two frequencies are equal.
The $\sigma_i$'s denote the Pauli matrices.
In order to treat the problem in its most general context, we will rather work
with the normalized variables $v_i=\omega_i/\omega_{max}$ where $v=(v_x,v_y)$ satisfies $\|v(t)\|\leq1$ for $t\in [0,T(v)]$  ($T(v)$ being the control duration) and $\Delta:=\omega/\omega_{max}$.
Such a transformation corresponds to a renormalization of the time by $\tau:=\frac{\omega_{max}}{2}t$, but by an abuse of notation we keep $t$ instead of $\tau$ in the remaining of the text.
The quantum Hamiltonian of the system now takes the form:
$$
\mathrm H = v_x(t)\sigma_x+v_y(t)\sigma_y+\Delta\sigma_z.
$$

The angular part of $v$ will be denoted $\mu(t)$ such that
\begin{equation}
\label{real control}
\begin{array}{l}
v_x(t)=v_0(t)\cos\mu(t),\\
v_y(t)=v_0(t)\sin\mu(t),\\
\end{array}
\end{equation}
$v_0$ being the amplitude of the control field. Note that the time dependance of most of the dynamical variables will be dropped throughout the paper in order to simplify the notation.

Writing the state of the system at time $t$ in the density matrix formalism, $\rho(t)$, the time evolution is given by the Von Neumann equation
$$
i\partial_t \rho(t)=[\mathrm H,\rho(t)],
$$
where one is working in a system of units such that $\hbar=1$. Since the system is closed, the states $\rho(t)$ are linked to the initial one $\rhoo$ by a unitary matrix $U(t)\in\SU(2)$ via the relation $\rho(t)=U(t)\rhoo U(t)^\dagger$. Note that $U(t)$ belongs to $\SU(2)$ and not U$(2)$ because the quantum Hamiltonian $\mathrm H(t)$ is an element of the Lie algebra $su(2)$.
The question of controlling the quantum system from a given initial state $\rho_0$ can then be translated to a control problem on $\SU(2)$, which is exactly the objective of this paper.
More precisely, one investigates the problem of finding the optimal control $v^*$ (the sign $*$ will indicate the optimal solutions in the rest of the paper) steering the system from the identity matrix $U(0)=I\in\SU(2)$ to a target state $U\f\in\SU(2)$ while minimizing the time $T(v)$ to get there.
We will use the simplified notation $t\f:=T(v\f)$.
The dynamics governing the system for this \dfn{time-optimal} control problem is given by the Schr\"odinger equation
\begin{equation}
\label{Dynamic U}
i\partial_t U(t)=\mathrm H(t)U(t).
\end{equation}
\subsection{Choice of coordinates}
Let us recall some characteristics of the group $\SU(2)$ we are going to work with.
Elements $U\in\SU(2)$ are the $2\times2$ matrices with complex entries satisfying $\det(U)=1$.
Defining $\i=\mn i\sigma_z$, $\j=\mn i\sigma_y$ and $\k=\mn i\sigma_x$, a possible parametrization of $U$ is
\begin{equation}
\label{U quaternion dec}
U
=
\left(\begin{array}{cc}
 x_1+\mn i {x_2} & {x_3}+\mn i {x_4} \\
 -{x_3}+\mn i {x_4} & {x_1}-\mn i {x_2}\\
\end{array}
\right) = {x_1}\id+{x_2}\i+{x_3}\j+{x_4}\k
\end{equation}
where the $x_i$'s are real.
This is the quaternion representation of $U$ and is related to the group of rotations $SO(3)$.
Indeed, there exists $\alpha\in[0,4\pi]$ and $\vv n\in S^2(0,1)$ (the unit sphere centered at the origin) such that
\begin{equation}
\label{quaternion}
\begin{array}{l}
x_1=\cos(\alpha/2)\\
x_2=\sin(\alpha/2)n_z\\
x_3=\sin(\alpha/2)n_y\\
x_4=\sin(\alpha/2)n_x\\
\end{array}
\end{equation}
and that $U$ represents a rotation of angle $\alpha$ about the unit axis $\vv n$.
However, since $\det(U)=1$, the $x_i$'s satisfy the relation $\sum_{i=1}^4x_i^2=1$ and the quaternion parametrization uses one parameter more than needed.
This remark partly motivates our choice of instead considering the so-called Hopf parametrization.
In term of the Hopf variables $\{\theta_1,\theta_2,\theta_3\}$, $U$ can be written as
\begin{equation}
\label{U-thetai decomposition}
U(\theta_1,\theta_2,\theta_3)
=
\left(\begin{array}{cc}
~\cos\theta_1e^{i\theta_2} &\sin\theta_1e^{i\theta_3} \\
-\sin\theta_1e^{-i\theta_3} & ~~\cos\theta_1e^{-i\theta_2}\\
\end{array}
\right)
\end{equation}
where the domain of definition will be defined below.
The main advantage of the Hopf parametrization is that Eq.~\eqref{Dynamic U} translates nicely in its variables, which give the simple form
\begin{equation}
\label{Dynamic theta_i det}
\left(\begin{array}{c}
\dot {\theta}_1\\
\dot {\theta}_2\\
\dot {\theta}_3\end{array}\right)
=
\left(
\begin{array}{c}
u_1\\
-\tan\theta_1~u_2-{\Delta}\\
~~\cot\theta_1~u_2-{\Delta}\\
\end{array}
\right)
\end{equation}
where the normalized "rotated" controls
\begin{equation}
\label{control change}
\begin{array}{l}
u_1=-v_0\sin(\mu+\theta_2+\theta_3),\\
u_2=-v_0\cos(\mu+\theta_2+\theta_3),\\
\end{array}
\end{equation}
have been used and satisfy $\|u\|\leq 1$.
The explicit calculations for obtaining equations Eq.~\eqref{Dynamic theta_i det} can be found in Appendix~\ref{app0}.
The control problem reads now as follows:
\textit{
Given a target state $(\thf_1,\thf_2,\thf_3)$, find the control $u\f$ steering  $\theta_i(0)\mapsto\theta_i\f$ while minimizing the control duration $T(u)$.}\\
Note that the condition $U(0)=I$ translates in terms of Hopf variables as
\begin{equation*}
\begin{array}{l}
\theta_1(0)=0,\\
\theta_2(0)=0,\\
\theta_3(0)=\text{undefined}.\\
\end{array}
\end{equation*}

In later parts of this paper, we will also make use of the Euler parametrization $U(\psi,\theta,\phi)$ which will allow us to easily visualize the rotations.
The Euler variables are related to the Hopf coordinates by
\begin{equation}
\label{HopfEuler}
\begin{array}{rcl}
\psi&=&\theta_2+\theta_3,\\
\theta&=&2\theta_1,\\
\phi&=&\theta_2-\theta_3.\\
\end{array}
\end{equation}
Recalling the close relate between elements of $\SU(2)$ and rotations $\rot(\psi,\theta,\phi)\in\SO(3)$, in terms of Euler coordinates, unitary matrices are given by  $U(\psi,\theta,\phi)=\e^{i\psi\frac{\sigma_z}{2}}\cdot\e^{i\theta\frac{\sigma_y}{2}}\cdot\e^{i\phi\frac{\sigma_z}{2}}$.
The Euler variables are taken to be in the domains $\phi\in[-\pi,\pi)$, $\psi\in[-2\pi,+2\pi)$ and $\theta\in[0,\pi]$.
The domains of the $\theta_i's$ variables can then be deduced directly from Eqs.~\eqref{HopfEuler}.
\section{The Pontryagin Maximum Principle}%
\label{sec3}
The problem of finding the time-optimal control for steering the system from an initial state to a fixed target can be decomposed into three steps, each of which brings its own considerations and particular methodology.
Firstly, one must find the optimal candidates for the control functions.
Here, we can have two types of controls, either regular or singular type.
The second step consists in computing the trajectories associated with the different controls. Finally, one has to determine the right trajectory which reaches the desired target state.
While the two first steps are treated through the Pontryagin Maximum Principle (PMP) \cite{pont}, depending on the optimal problem considered the last step may require numerical methods.

The present section is intended to expose the main features of the PMP and to apply it to our particular problem.
\subsection{Theory}
\label{sec 3A}
This section aims at presenting the elements of the theory of PMP used to solve our control  problem.
In order to be as pedagogical as possible, the general equations arising from the theory \cite{pont,bonnardbook} will be followed by their translation in our particular context.

Consider a controlled system
\begin{equation}
\label{general dynamic}
\dot{x}= f(t,x(t),u(t))
\end{equation}
where $x(t)=(x_1,x_2,\cdots,x_n)\in\R^n$, $u(t)\in\Omega\subset\R^m$ for $t\in[0,T]$ where $T:=T(u)$ as before.
Let $x_u(\cdot)$ denotes the trajectory associated to the control $u$.
Given two points $x_0,x^*\in\R^n$, one aims at finding $u$ such that $x_{u}(0)=x_0$, $x_{u}(T)=x^*$ and minimizing (or eventually optimizing) the cost functionnal
\begin{equation}
\label{cost function}
c(T,u)=g(T,x_u(T))+\int_0^T f^0(t,x_u(t),{u}(t))~dt.
\end{equation}
The function $f^0$ is the running cost which depends on the whole trajectory $x_u(\cdot)$ whereas $g$ is the final cost which depends only on the final time and state.
The \dfn{Hamiltonian} of the system driven by $u$ is also called PMP \textit{pseudo-Hamiltonian} and is denoted
\begin{equation}
\label{pseudo-Hamiltonian}
H(t,x,p,\p^0,u)= p\cdot x+\pf.
\end{equation}
The term $p\cdot x$ denotes the scalar product between the state vector $x$ and a vector $p(t)\in\R^{n}$ called the \dfn{adjoint state}, which is required to be continuous all along an optimal trajectory.
The constant $\p^0$ is taken to be negative (for a maximum principle) and should be such that $\p^0$ and $p$ never simultaneously vanish.
Note that the overall vector $(p,\p^0)\in\R^{n+1}$ is defined up to a positive constant factor.

At this stage, we express our problem in this formalism.
First, Eq.~\eqref{general dynamic} corresponds to our system given in Eq.~\eqref{Dynamic theta_i det} with $x_i:=\theta_i$.
The norm of the controls $u(t)$ are bounded to $\|u\|\leq1$.
Since the goal is to minimize the control time, the cost function is $c(T,u)=T$ and one can consider $g\equiv0$ and $f^0\equiv1$ in Eq.~\eqref{cost function}.
Finally, after factorizing the two control components, the PMP pseudo-Hamiltonian takes the form
\begin{equation}
\label{H det}
\begin{array}{l}
H(t, \vec\theta, p,\p^0, u)=\\
u_1p_1+u_2(-p_2\tan{\theta_1}+p_3\cot{\theta_1})-(p_2+p_3)\Delta+\p^0
\end{array}
\end{equation}
where $\vec \theta:=(\theta_1,\theta_2,\theta_3)$.
The PMP suggests that a \textit{necessary} condition for a control $u\f(t)$ to optimize (minimize in our case) the cost functional is to maximize the PMP pseudo-Hamiltonian $H(t)$ at any time.
This leads to the PMP Hamiltonian $\H_{u\f}(t,x, p,\p^0)$, denoted with a script letter, which is free of an explicit control dependence and which satisfies
\begin{equation*}
\label{PMP1}
\H_{u\f}(t,x, p,\p^0)=\max_{u\in \Omega} H(t, x,p,\p^0, u), \quad \forall~t \in [0,T].
\end{equation*}
Due to the fact that $u_1$ and $u_2$ can be factorized in the pseudo-Hamiltonian $H$ (see~Eq. \eqref{H det}), we find that the controls are given by
\begin{equation}
\label{controls}
\begin{array}{l}
\vspace{1mm}
u_1(t)=\frac{p_1}{N}, \\
u_2(t)=\frac{-p_2\tan\theta_1+p_3\cot\theta_1}{N},\\
\end{array}
\end{equation}
where
\begin{equation}
\label{N value}
N=\sqrt{p_1^2+[-p_2\tan\theta_1+p_3\cot\theta_1]^2}
\end{equation}
is such that $u_1^2+u_2^2=1$ \cite{3level,bonnardbook}.
Such controls, which are well defined when $N\neq 0$ only, are said to be \dfn{normal} and the associated trajectories $\vec\theta(\cdot)$ are called \dfn{regular}.
When $N=0$, computational analysis (detailed below) reveals that the controls, said to be \textit{singular}, must vanish.
The associated trajectories are also called \dfn{singular} and are non-trivial only when $\Delta\neq 0$.
However, for both $\Delta=0$ and $\Delta\neq0$, we can show that singular trajectories as well as mixtures of regular and singular trajectories are \textit{never} optimal \cite{newmath}.
In other words, the time-optimal controls for our problem are necessarily normal and consequently, we will restrict our study to these ones only.
In addition, the PMP states that the final PMP Hamiltonian must satisfy
\begin{align}
\H_{u\f}(T)&= -\p^0\frac{\partial g}{\partial t}(T),\label{transversality} \\
\frac{\partial\H_{u\f}}{\partial t} &=\frac{d\H_{u\f}}{dt}\label{H no time},
\end{align}
for almost all $t\in [0,T]$.
Equation~\eqref{transversality} is called the \dfn{transversality} condition and takes the above form when we apply the PMP formalism to our problem for some non-fixed final time $T$.
Equation~\eqref{H no time} is derived from the fact that the PMP pseudo-Hamiltonian does not depend explicitly on time.
Note that in our case, since $g\equiv 0$, those two relations merge into the relation
\begin{equation}
\label{H=0}
\H_{u\f}(\vec\theta,p,\p^0)= 0.
\end{equation}
In the following, the constant $\p^0$ is normalized to -1.
Finally, the PMP ensures that the dynamics along an optimal trajectory is given by the Hamilton set of $2n$ differential equations

\begin{align}
\frac{\partial \H_{u\f}}{\partial p} &=\dot{x},\nonumber \\
-\frac{\partial \H_{u\f}}{\partial x}&=\dot{ p}.\nonumber
\end{align}

When normal controls are considered, this system takes the form
\begin{equation}
\label{Dynamic regular}
\left\{
\begin{array}{l}
{\left(\begin{array}{c}
\dot {\theta}_1\\
\dot {\theta}_2\\
\dot {\theta}_3\end{array}\right)
=
\left(
\begin{array}{c}
p_1\\
\p_2\tan^2\theta_1-{\Delta}\\
-\p_2-{\Delta}\\
\end{array}
\right)}\\
{\left(\begin{array}{c}
\dot {p}_1\\
\dot {p}_2\\
\dot {p}_3\end{array}\right)
=
\left(
\begin{array}{c}
-\p_2^2\tan\theta_1\sec^2\theta_1\\
0\\
0\\
\end{array}
\right)}\\
\end{array}
\right.
\end{equation}
where the adjoint variables have been normalized to $p_i':=\frac{p_i}{N}$, the prime has been dropped to simplify the notations and we used the constants of the motion
\begin{subequations}
\begin{align}
p_2(t)&\equiv \p_2,\nonumber\\
p_3(t)&\equiv 0,\nonumber
\end{align}
\end{subequations}
to simplify the equations.
The relation $p_3(t)\equiv 0$ is deduced from the expression of $u_2(0)$ in Eqs.~\eqref{controls}.
Since $\theta_1(0)=0$, the $p_3\cot\theta_1$ term would be infinite in $t=0$ if $p_3(0)\neq0$, which can not be the case since the controls have finite intensity. Using this condition, it is also straightforward to check that singular controls are null.
The term $N(t)$ is equal to zero on a given time interval only if $p_1(t)=0$.
The associated singular trajectories correspond to a freely evolving system whose dynamics is governed by the detuning term $\Delta$.

\section{Optimal trajectories without detuning}
\label{sec4}
We recall that the case without detuning corresponds to $\Delta=0$. As already mentioned, the only controls to consider are the normal ones.
In particular, since such controls satisfy $\|u\|=1$, the controls are uniquely characterized (modulo $2\pi$) by the angular parameter $\beta(t):=\mu(t)+\psi(t)$ (see Eqs.~\eqref{control change}).

In this section, we explicitly write the solutions of the system of dynamical equations~\eqref{Dynamic regular} in terms of the Euler angles parametrization $(\psi,\theta,\phi)$.
The full problem is solved for some typical targets frequently encountered in quantum computing \cite{nielsen}.

\subsection{The general case}%

Using the Eqs.~\eqref{HopfEuler} and setting $\Delta=0$, the dynamics satisfied by the regular extremals can be written in terms of the Euler parameters as
\begin{equation}
\label{Dynamic Euler}
\begin{array}{l}
{\left(\begin{array}{c}
\dot {\psi}\\
\dot {\theta}\\
\dot {\phi}
\end{array}\right)
=
\left(
\begin{array}{c}
\p_2(\tan^2\theta_1-1)\\
2p_1\\
\p_2\sec^2\theta_1\\
\end{array}
\right).}\\
\end{array}
\end{equation}
Using the two definitions of the controls given in Eqs.~\eqref{control change} and Eqs.~\eqref{controls} as well as the dynamics described by the Eq.~\eqref{Dynamic Euler}, we get two relations between the variables $\psi(t)$, $\phi(t)$ and $\mu(t)$:
\begin{equation}
\label{cst motion}
\dot\mu
=2\p_2
\quad\text{and}\quad
\dot\mu+\dot{\psi}-\dot\phi
= 0.
\end{equation}
Since $\p_2$ is constant, the first of these two equations implies that the real angular control is a linear function of the time
$$\mu(t)=\mu(0)+2\p_2t.$$
We can show that $\beta(0)=-\frac{\pi}{2}$ \cite{newmath} and from this relation together with the fact that $\phi(0)=-\psi(0)$ (since $\theta_2(0)=0$), we directly compute $\mu(0)=\phi(0)-\frac\pi2$ by using the definition of $\beta(t)$.

Equations \eqref{cst motion} also provide us the main tool needed to visualize the extremals as their projection on the sphere $S^2$.
Indeed, given an extremal $(\psi(t),\theta(t),\phi(t))$ associated to the regular control $u(t)$ for $t\in[0,T(u)]$, let $\gamma(t):=(\theta(t),\phi(t))$ be the projection of this extremal on the sphere. The variable $\theta(t)$ gives the vertical inclination and $\phi(t)$ the azimuthal angle with respect to the $x$- axis.
We assume that $\phi(0)$ and $\p_2$, which define a particular trajectory, are known.
We then have the following property (see Appendix~\ref{app1} for the proof)  which has been also established in Ref. \cite{boozer}.
\begin{proposition}
\label{trajectory circle}
Let $\bar{\theta}=\arctan(\frac{1}{\p_2})$ and $\bar{\phi}=\phi(0)+\frac{\pi}{2}$.
The projected trajectory $\gamma(t)=(\theta(t),\phi(t))$ defines a circle around the fixed axis $\vec{n}=(\bar{\theta},\bar{\phi})$.
Moreover, $\gamma(t)$ is traveled with constant speed $\|\dot\gamma(t)\|=2$.
The final time $\tf$ is equal to $\frac12$ the arc length of $\gamma(t)$.
\end{proposition}
Notice that the factor $+\frac\pi2$ appearing in the definition of $\bar\phi$ comes from the relation $\beta(0)=-\frac\pi2$ previously established.
\begin{figure}[ht]
\includegraphics[width=0.7\textwidth]{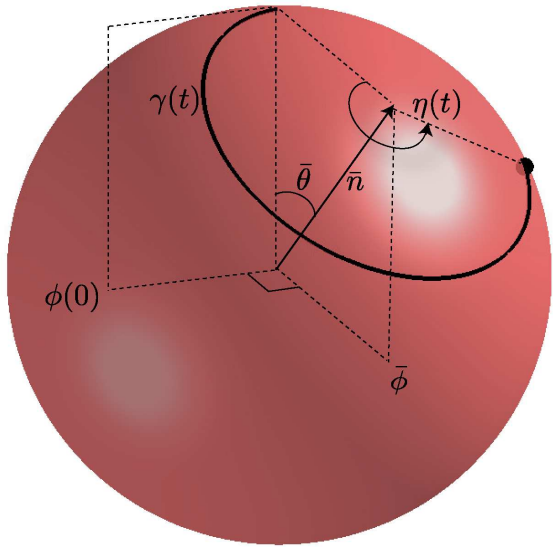}
\caption{Illustration of the different angular parameters used in Prop.~\ref{trajectory circle} to characterize a projected trajectory $\gamma(t)$.}
\label{sphere parameters}
\end{figure}

Examples of such projected trajectories are illustrated in Fig.~\ref{y-z traj}.
In the situation where $\p_2=0$, Eqs.~\eqref{Dynamic regular} are directly integrable and hence we obtain that $\gamma(t)$ describes a great circle, which is the limit case of the previous Proposition~\ref{trajectory circle}.

Using Eqs. \eqref{cst motion}, we deduce the explicit expression for the time evolution of Euler parameters
\begin{equation}
\label{Euler trajectory}
\begin{array}{l}
\theta(t)=\cos^{-1}(1+\sin^2\bar\theta(\cos\eta(t)-1)),\\
\phi(t)=\phi(0)+\text{sign}(\p_2)\frac{\pi}{2}+\tan^{-1}\left(\frac{\sin\eta(t)}{\cos\bar\theta(\cos\eta(t)-1)}\right),\\
\psi(t)=-2\phi(0)+\phi-2\p_2t,\\
\end{array}
\end{equation}
where $\eta(t):=\frac{2t}{\sin\bar\theta}$ is the angular parametrization of the circle section drawn by $\gamma(t)$ around its axis $\bar n$ (see Fig.~\ref{sphere parameters}).
In particular, the time $t$ for which $\gamma(t)$ traces out a complete circle on the sphere corresponds to an angle of $\eta(t)=\sgn(\p_2)2\pi$.
In the limiting case where $\p_2=0$, the equation for $\phi(t)$ given in Eqs.~\eqref{Euler trajectory} takes the simple form $\phi(t)\equiv\phi(0)$.
\subsection{Time-optimal controls}

We can show that all unitary matrices $U\f\in\SU(2)$ are reached by a unique regular control $u:[0,\tf]\mapsto \SU(2)$ with $|\eta(\tf)|<2\pi$, for all unitary matrices which do not correspond to $z$-rotations.
Moreover, these controls are precisely the time-optimal ones.
The condition $|\eta(\tf)|<2\pi$ implies that time-optimal trajectories necessarily trace out circular arcs of angle $|\eta(\tf)|<2\pi$ and hence, they turn less than once around their axis $\bar n$.
All regular controls satisfying $|\eta(T)|<2\pi$ ($T=T(u)$) are then uniquely denoted $u_{U}$ where $U\in\SU(2)$ is the target unitary they optimally generate.
The case $|\eta(T)|=2\pi$ corresponds to rotations along the $z$-axis and, in this case, all the trajectories corresponding to the same inclination angle $\bar\theta$ will generate the same unitary matrix in the same time.
This special case will be discussed further in the next section.

To completely solve the problem at hand, it remains to find the explicit parameters  $\p_2$ and $\phi(0)$ (defining any regular controls up to the final time) which steer the system to the desired target.
In fact, work done in \cite{newmath} shows that
\begin{equation}
\label{p2}
\p_2=\frac{\sin(\phi\f-\phi(0))}{\tan(\frac{\theta\f}{2})},
\end{equation}
and hence we need to only consider the single parameter $\phi(0)$.
Consequently, Eqs.~\eqref{Euler trajectory} is a system of three equations of two variables $\phi(0)$ and $\tf$ and the exact solution for a general $U\f=(\psi\f,\theta\f,\phi\f)$ can be found numerically.

Some cases are of particular practical interests, namely the unitary matrices describing rotations about the $z$- axis or about any axis in the $(x,y)$-plane.
The full solution for each of these two classes of targets are presented in the next sections, and we will see that the initial parameter $\phi(0)$ can be found analytically in both cases.
Interestingly, an analytical formula for the optimal time $\tf$ is found as a function of the target $U\f$.
The general case will finally be briefly discussed.

\subsection{Rotation about the $z$- axis}
\label{z- axis}
\begin{figure}[ht]
\includegraphics[width=0.7\textwidth]{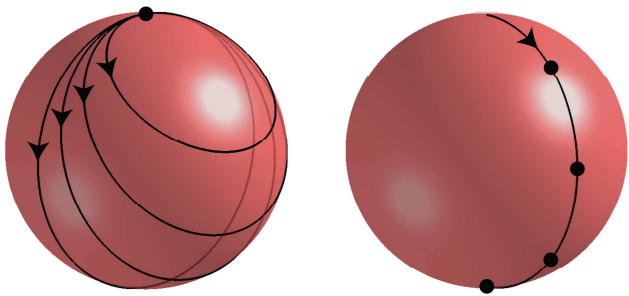}
\caption{On the left, the projected trajectories for the optimal synthesis of $U\f=e^{i\lambda\frac{\sigma_z}{2}}$ with $\lambda=\frac{\pi}{2},\pi,\frac{3\pi}{2},2\pi$.
On the right, the projection of the optimal trajectories of $U\f=e^{-ia\frac{\sigma_z}{2}}e^{ib\frac{\sigma_y}{2}}e^{ia\frac{\sigma_z}{2}}$ with $(a,b)=(0,\frac{k\pi}4)$ where $k=1,2,3,4$. The dots indicate the position of the target states.}
\label{y-z traj}
\end{figure}

Consider the problem of reaching a target of the form $U\f=e^{i\lambda\f\frac{\sigma_z}{2}}$ where $\lambda\f\in[-2\pi,2\pi]$.
The final state is characterized by any pair $(\lambda\f,\phif)$, i.e.
$(\psi\f,\theta\f,\phi\f)=(\lambda\f-\phi\f,0,\phif)$.
Since the definition of $U\f$ depends only on $\lambda\f$, the time-optimal trajectories associated to any value of $\phi\f$ should not influence the final time $\tf$ taken to reach the target.
Note that the projected trajectory $\gamma(t):=(\theta(t),\phi(t))$ starts and ends at the North Pole of the sphere since $\theta(0)=\theta\f=0$.
As a consequence of this remark together with Proposition~\ref{trajectory circle},  $\gamma:[0,\tf]\rightarrow S^2$ describes a complete circle and direct computation shows that the initial parameters leading to the target are given by\begin{equation}
\label{parameters z- axis}
\begin{array}{rcl}
\p_2     &=& \sgn(\lambda\f)\cot\left(\cos^{-1}\left(1-\frac{|\lambda\f|}{2\pi}\right)\right)\\
         &\text{and}&\\
\tf      &=& \frac12\sqrt{4\pi|\lambda\f|-|\lambda\f|^2}\\
\end{array}
\end{equation}
from which we can explicitly write the time-optimal trajectory given by Eqs.~\eqref{Euler trajectory}.
Examples of projected trajectories for four different $z$-rotation unitary matrices are shown on the left sphere of Fig.~\ref{y-z traj}. Note that equations similar to Eqs.~\eqref{parameters z- axis} have been encountered in \cite{khaneja1} for the time-optimal control on three coupled spins.

\subsection{Rotation in the $(x,y)$- plane}
\label{xy- plane}
Now, we consider a target corresponding to a rotation in the $(x,y)$- plane
given by $U\f=e^{i(-a)\frac{\sigma_z}{2}}\cdot e^{ib\frac{\sigma_y}{2}} \cdot e^{ia\frac{\sigma_z}{2}}$ which is defined by the Euler variables $(\psi\f,\theta\f,\phi\f)=(-a,b,a)$ with $b\in[0,2\pi)$.
This special choice for the domain $[0,2\pi)$ of $b$ will find its motivation later when we will consider the opposite rotation $-U$.
At this point, a remark about the behaviour of the Hopf variable $\theta_2$ should be pointed out.
We have observed that $\theta_2(t)$ is monotonic along a time-optimal trajectory, meaning that it must either strictly increase or strictly decrease otherwise it is equal to zero all along the trajectory \cite{newmath}.
In the present case, since $\theta_2(0)=\theta_2\f=0$, the third situation applies which implies that $\dot\theta_2=0$ and $\theta_2(t)=0$ along the complete trajectory.
By comparing this equality with Eqs.~\eqref{Dynamic regular}, it becomes clear that we are in the very special case where $\p_2=0$ and that the only variable among Hopf angles and adjoint state variables which \textit{is not} constant is $\theta_1$.
All the equations among Eqs.~\eqref{Dynamic Euler} are directly integrable, and hence we obtain that $\gamma(t)$ must follow a great circle.
The initial conditions and the final time for reaching the target can then be written as
\begin{equation}
\label{parameters xy- plane}
\begin{array}{rcl}
\p_2 = 0,~~\phi(0)  = a ~~~\text{and}~~~ \tf = \frac b2.\\
\end{array}
\end{equation}
Examples of projected trajectories for four different unitary matrices corresponding to rotations in the transverse plane are shown on the right sphere of Fig.~\ref{y-z traj}.

\subsection{Optimal control in SO(3)}

In the context of many experiments (in NMR or quantum gate generation for instance), the global phase of a unitary matrix $U$ is not relevant, in the sense that the action of two opposite evolution operators $U$ and $-U$ on an identical initial state $\rho_0$ results in identical states which are experimentally undistinguishable.
The global phase issue has been discussed in \cite{kobzar,JanichSchulte,schulteSporl}. In Ref. \cite{schulteSporl}, the authors point out that the time to optimally generate $U$ or $-U$ generally differ.
The control problem on $\SU(2)$ in this experimental context then translates as a control problem on the group of rotations $\SO(3)$.

The time-optimal control problem on $\SO(3)$ can be reformulated as follows:
\textit{Given two opposite unitary matrices $U$ and $-U$, which one of $U$ or $-U$ is the fastest to generate in the context of time-optimal control on $\SU(2)$?}
If $U\f\in\{U,-U\}$ denotes the answer to this question, then the time optimal control $u(t)$ for generating $U\f$ also corresponds to the time-optimal control for generating the associated rotation $\rot\f\in\SO(3)$.
In the following, we aim at answering the above question for any pair of unitary matrices $U$ and $-U$.
After considering the two classes of unitary matrices studied in Sec.~\ref{z- axis} and \ref{xy- plane}, we will finally find the class of $U\in\SU(2)$ such that $U$ and $-U$ are optimally reached in the same time.
In order to simplify the notations, the variables related to $-U$ will be denoted with the symbol $\til \cdot$.

{\flushleft\textbf{\underline{Rotation about the $z$- axis.}}}\\
Let $U=e^{i\lambda\frac{\sigma_z}{2} }$ as before and $-U=e^{i\til{\lambda}\frac{\sigma_z}{2} }$.
The parameter $\til\lambda=\lambda-\sgn(\lambda)\cdot2\pi$ is chosen to be in the domain $\til\lambda\in[-2\pi,2\pi]$.
Note that $\lambda$ and $\til\lambda$ have opposite signs.
Using the equations for the final time given in Eq. \eqref{parameters z- axis}, we deduce that:

\begin{proposition}
The rotation $\rot\f=\rot(\lambda-\phif,0,\phi\f)$ is optimally generated by $U$ if $|\lambda|<\pi$ and by $-U$ otherwise.
The time for generating $U$ and $-U$ are the same when $|\lambda|=\pi$.
\end{proposition}

\begin{figure}[ht]
\includegraphics[width=0.7\textwidth]{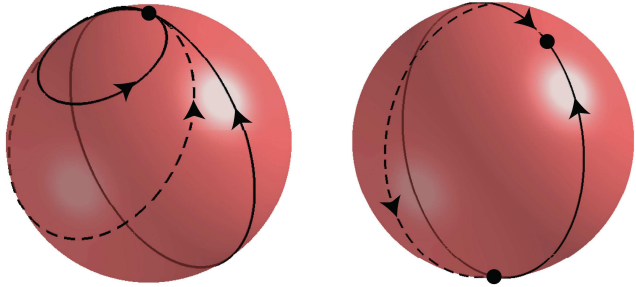}

\caption{On both figures, the two full lines correspond to the trajectories for $U\f$ and $-U\f$, where the dots correspond to the end points of the trajectories and the arrows to the travel direction.
On the left, we see the projected trajectory of a $z$-rotation where $U\f=e^{i\frac{\pi}{4}\frac{\sigma_z}2}$.
On the right, the figure displays the projected trajectory of a rotation on the $(x,y)$-plane where  $U\f=e^{i\frac{\pi}{5}\frac{\sigma_y}2}$.
The dashed lines are the trajectories for the limit case $U$ such that $\tf(U)=\tf(-U)$.}
\end{figure}

{\flushleft\textbf{\underline{Rotation in the $(x,y)$- plane.}}}\\
Let $U=e^{-ia\frac{\sigma_z}{2}}e^{ib\frac{\sigma_y}{2}}e^{ia\frac{\sigma_z}{2}}$ be a unitary matrix where the domains of definition are $a\in[-\pi,\pi]$ and $b\in(0,2\pi)$.
Let $- U=e^{-i\til a\frac{\sigma_z}{2}}e^{i\til b\frac{\sigma_y}{2}}e^{i\til a\frac{\sigma_z}{2}}$ be the unitary matrix opposite to $U$.
In order to remain in the prescribed domain, $- U$ is characterized by $\til a=a\pm\pi$ and $\til b=2\pi-b$ where the sign $\pm$ is chosen such that $a$ and $\til a$ have opposite signs.
Using the equation for the final time given in Eqs.~\eqref{parameters xy- plane}, we deduce that:

\begin{proposition}
\label{same time U -U p2=0}
The rotation $\rot\f:=\rot(-a,b,a)$ is optimally generated by $U$ if $b<\pi$ and by $-U$ if $b>\pi$.
The time for generating $U$ and $-U$ are the same when $b=\pi$.
\end{proposition}

In other words, for a rotation about any axis in the transverse plane, $\tf=\til \tf$ if and only if $U$ describes a $\pi$-rotation about this axis.

{\flushleft\textbf{\underline{General case.}}}\\
In order to find which one of $U$ and $-U$ is the fastest to generate in the most general case, we will proceed as before by first finding the class of unitary matrices satisfying $t\f=\til t\f$.
We have already seen that $\pm\pi$-rotations about the $z$- axis or any axis in the transverse plane are such that the associated $U$ and $-U$ matrices are optimally reached in the same time.
Let us now consider a general rotation $\rot(\psi\f,\theta\f,\phi\f)$ about any axis which is not included in the two families already treated and let $U$ and $-U$ be the two unitary matrices generating this rotation.
We aim at finding the criterion that should be filled by $\rot$ in order to have $\tf=\til \tf$.

First, we know from Proposition~\ref{trajectory circle} that the final time $\tf$ for reaching $U$ is equal to half of the length of the projected trajectory $\gamma$.
Consequently, since $\tf=\til \tf$, $\gamma$ and $\til\gamma$ have the same length.
Coming back to the Hopf parametrization, and writing the parameters for $-U$ as a function of $U$, we get:
\begin{equation}
\label{U -U Euler}
\begin{array}{l}
\til{\thf_1}=\thf_1 ~~~\text{and}~~~\til{\thf_2}=\thf_2-\sgn(\thf_2) \pi .\\
\end{array}
\end{equation}
In particular, the Euler variables $\theta\f$ and $\til\theta\f$ are equal.
Summarizing what has been deduced so far, $\gamma$ and $\til\gamma$ have the same length, end up at the same inclination angle $\theta\f$ while describing a circle on the sphere starting at the North Pole.
The only possibility is that $\gamma$ and $\til\gamma$ describe a circle of identical radius, but one clockwise and the other anti-clockwise since $\theta\f_2$ and $\til\theta\f_2$ have opposite signs.
In fact, we can deduce that $\theta_2\f=-\til\theta_2\f$ and inserting this relation in Eq. \eqref{U -U Euler}, the condition for $U$ such that $\tf=\til\tf$ is
\begin{equation}
\label{theta2=pi/2}
\thf_2=\pm\frac\pi2.
\end{equation}
In other words, the unitary matrices corresponding to $\pi$-rotations ($\rot$)  about \textit{any} axis are the ones which take as much time to generate as their opposite.
This can be seen using the quaternion parametrization introduced in Sec.~\ref{sec2}.
Indeed, $\theta_2\f=\pm\frac\pi2$ implies that $x_1=0$ in Eq. \eqref{U-thetai decomposition}.
In terms of quaternions, this leads to $x_1=\cos(\alpha/2)$ from which we deduce that $\alpha=\pm\pi$.
Recalling that $\alpha$ denotes the rotation angle of $U$ about a given axis,
the rotation associated to $\theta\f_2=\pm\frac\pi2$ is a $\pm\pi$- rotation about this axis (see Fig.~\ref{TvsrotAngle}).
Finally, we deduce:
\begin{proposition}
\label{same time U -U general}
Let $U\in\SU(2)$ and $\theta_2\f$ be the corresponding Hopf parameter.
Then $\tf<\til\tf \iff |\theta_2\f|<\frac{\pi}{2}$.
In particular, $\tf=\til\tf \iff |\theta_2|=\frac{\pi}{2}$.
\end{proposition}

The initial value  $\phi(0)$ can be found numerically by inverting the system of equations \eqref{Euler trajectory} at the target state $(\psi\f,\theta\f,\phi\f)$.
Once $\phi(0)$ is known, the time-optimal trajectory is described by the Eqs.~\eqref{Euler trajectory} using the definition of the adjoint variable $\p_2$ given by Eq.~\eqref{p2}.
Once these two variables known, the final time $\tf$ can finally be found by inverting the system of equations~\eqref{Euler trajectory}.

Figure~\ref{TvsrotAngle} shows the optimal time to generate unitary matrices defined according to their quaternion's definition given in Eq.~\eqref{quaternion}.
Three axes $\vec n_y, \frac{1}{\sqrt 2}(\vec n_y+\vec n_z)$ and $\vec n_z$ are considered, corresponding to an axis progressively tilted from the $(x,y)$-plane to the $z$-axis.
Note the monotonic evolution of the time function on the intervals $[0,2\pi]$ and $[2\pi,4\pi]$ in Fig.~\ref{TvsrotAngle}.
As expected, the matrices satisfying $\tf=\til\tf$ (intersection of the solid and dashed lines) are exactly the ones for which $\alpha=\pi+k\cdot 2\pi$.
At these points (black dots), the corresponding rotations for $\vec n =\vec n_y$ and $\vec n=\frac{1}{\sqrt 2}(\vec n_y+\vec n_z)$ are the well-known refocusing \cite{spin} and Hadamard gates \cite{nielsen}, respectively.
\begin{figure}[ht]
\label{TvsrotAngle}
\centering
\includegraphics[width=0.7\textwidth]{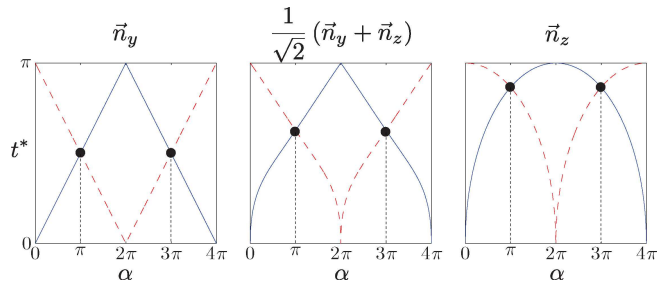}
\caption{
Plot of the minimum time $\tf$ to generate the rotations around the $\vec n_y, \frac{1}{\sqrt 2}(\vec n_y+\vec n_z)$ and $\vec n_z$ axes as a function of the rotation angle $\alpha$.
The blue (solid) and red (dashed) curves represent the time to generate $U$ and $-U$ respectively.
The black dots and vertical dotted lines underline the angles $\alpha$ for which $\tf=\til\tf$.
These angles are $\alpha =\pi,3\pi$, as expected  (see Eq.~\eqref{theta2=pi/2}).
}
\label{TvsrotAngle}
\end{figure}

\section{Optimal trajectories with detuning}
\label{sec5}

In this section we consider the case where the frequency of the transverse magnetic field $\vec \B_1$ is not on resonance with the Larmor frequency.
In the rotating frame, the system is subject to a $z$-magnetic field proportional to $\Delta\neq0$.
In this case, both regular and singular controls lead to non-trivial trajectories but, as already discussed in Sec.~\ref{sec 3A}, the time-optimal ones are necessarily regular, that is, time-optimal controls are never vanishing.
In the following, we discuss the shape of the regular trajectories and the set of the time-optimal ones is given as a function of $\Delta$.
Notice that the analysis is completely general and applies to any possible value of $\Delta$, as long as the rotating wave approximation is valid.
Finally, for any $U\in\SU(2)$, we will tackle the problem of finding which one among $U$ and $-U$ is the fastest to generate for a given value of $\Delta\in\R$, in the context of control problems in $\SO(3)$.

\subsection{The general case}
\label{sec5A}

Let $\Delta\neq0$ and consider the dynamics of the system given by Eqs.~
\eqref{Dynamic theta_i det} in terms of the Euler parameters
\begin{equation*}
\left(\begin{array}{c}
\dot \psi\\
\dot {\theta}\\
\dot \phi\end{array}\right)
=
\left(
\begin{array}{c}
u_2(-\tan\theta_1+\cot\theta_1)-2\Delta\\
2u_1\\
u_2(-\tan\theta_1-\cot\theta_1)~~~~~~~\\
\end{array}
\right).
\end{equation*}
Replacing the control values (see Eqs.~\eqref{controls}) in these equations, the dynamics of the system governed by the associated regular control is then
\begin{equation}
\label{Dynamic Euler det}
\begin{array}{l}
{\left(\begin{array}{c}
\dot {\psi}\\
\dot {\theta}\\
\dot {\phi}
\end{array}\right)
=
\left(
\begin{array}{c}
\p_2(\tan^2\theta_1-1)-2\Delta\\
2p_1\\
\p_2\sec^2\theta_1\\
\end{array}
\right).}\\
\end{array}
\end{equation}

Equations \eqref{cst motion} generalize in a straightforward way to the case with a detuning term as follows:
\begin{equation}
\label{cst motion det}
\dot\mu
=2\p_2-2\Delta
\quad\text{and}\quad
\dot\mu+\dot\psi-\dot{\phi}
= 0.
\end{equation}
\noindent

More generally, when comparing dynamics of Eqs.~\eqref{Dynamic Euler det} with the one without detuning given by Eqs.~\eqref{Dynamic Euler}, we notice that the only variable being influenced by the detuning $\Delta$ is $\psi(t)$.
In particular, the projected trajectories $\gamma(t)=(\phi(t),\theta(t))$ are of the same shape as the ones without detuning, i.e. they describe circles on the sphere.
More precisely, Proposition~\ref{trajectory circle} still holds with the same definition for the rotation axis $\bar n=(\bar\theta,\bar\phi)$.

%

\subsection{Time-optimal controls}
\label{sec5B}

In order to be as general as possible and include rotations about the $z$-axis in our study, every unitary matrix $U\f=e^{i\lambda\f\frac{\sigma_z}{2}}$ corresponding to a $z$-rotation will be characterized by $\phi\f=0$, that is $U\f=(\psi\f,\theta\f,\phi\f)=(\lambda\f,0,0)$.
To avoid confusion, targets  $\U=(\Psi,\Theta,\Phi)$ and $U\f=(\psi\f,\theta\f,\phi\f)$  will correspond to targets which are reached without and with detuning respectively.

As opposed to the case without detuning, when $\Delta\neq 0$, there may be more than one regular control $u$ reaching a target $U\f\in\SU(2)$ with $|\eta(T(u))|\leq 2\pi$.
We are then interested in finding which regular controls correspond to the time-optimal ones.
Recall that the regular controls are uniquely denoted $u_{\U}$ where $\U\in\SU(2)$ is the unitary matrix being generated by the regular control $u_{\U}$ \textit{without detuning}.
We consider the well-known end-point mapping \cite{bonnardbook}:
\begin{equation}
\label{end-point mapping}
\End\d:u_{\U}\mapsto \left(\Psi-2\Delta T(u_{\U}),\Theta,\Phi\right)
\end{equation}
mapping any regular control on its target in the presence of a detuning term $\Delta$.
We verify directly that when there is no detuning, this function maps a control $u_{\U}$ on $\U$, that is $\End_{0}(u_{\U})=\U$.
We can also check the validity of the mapping by noting 1) that the dynamical variables $\theta(t)$ and $\phi(t)$ are unaffected by the detuning and 2) that $\psi\f=\Psi-2\Delta T(u_{\U})$ by simply comparing the equations for $\dot\psi$ given in Eqs.~\eqref{Dynamic Euler} and \eqref{Dynamic Euler det}.

Clearly, the regular controls $u_{\U}$ reaching a desired target $U\f=(\psif,\theta\f,\phi\f)$ are  the ones for which $\Theta=\theta\f$, $\Phi=\phi\f$ and $\Psi-2\Delta T(u_{\U})=\psi\f\mod 4\pi$.
Since the two first relations are trivial to satisfy, we can then focus on the variable $\Psi$ only and consider the restriction (denoted $f\d$) of the end-point mapping $\End\d$ to the controls $u_{\U}$ satisfying $\Theta=\theta\f$ and $\Phi=\phi\f$.
These controls are then uniquely denoted $u_{\Psi}$.
The function $f\d$ takes the form:
\begin{equation}
\label{end-point mapping}
f\d:u_{\Psi}\mapsto \Psi-2\Delta T(u_{\Psi}).
\end{equation}
As expected, we have $f_0(u_{\Psi})=\Psi$ and the end-point mapping is given by $\End\d(u_{\Psi})=(f\d(u_{\Psi}),\Theta,\Phi)$.

With these tools available, we obtain the set of controls $u_{\Psi}$ corresponding to time-optimal controls under a detuning term $\Delta\neq 0$.
This set, denoted $\Omega\d^{(\theta\f,\phi\f)}$ or simply $\Omega\d$ (for $\theta\f$ and $\phi\f$ fixed), is given in the following Proposition.
\begin{proposition}
\label{optimal domain}
Let $\theta\f$ and $\phi\f$ being fixed.
If $|{\Delta}|\leq|\tan(\frac{\theta\f}{2})|$, then $\Omega\d=[u_{-2\pi},u_{2\pi}]$.
Otherwise, $\Omega\d=\left[u_{\Psi\bl},f\d^{-1}\left(f\d(u_{\Psi\bl})\pm4\pi\right)\right]$ where $\pm$ is the sign of $-\Delta$.
\end{proposition}
Here, $[u_{\Psi_{min}},u_{\Psi_{max}}]:=\{u_{\Psi}| \Psi\in[\Psi_{min},\Psi_{max}]\}$.
In particular, the time-optimal domain $\{u_{\Psi}\}\in\Omega_\Delta$ is such that $\Psi\in[\Psi_{min},\Psi_{max}]$.
The angle $\Psi\bl$ is chosen such that the adjoint variable $\p_2$ characterizing $u_{\Psi\bl}$ 
 satisfies $\p_2=\frac{1}{\Delta}$ and depends on $\Delta$.
The interested reader will find an idea of the proof in the Appendix \ref{app}.

\begin{figure}[ht]
\includegraphics[width=0.7\textwidth]{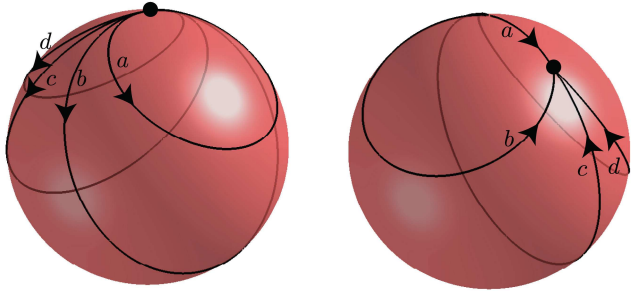}
\caption{Projected trajectories for the optimal synthesis of $U\f\in\SU(2)$ subjected to a detuning field of intensity $\Delta=0,\frac{1}{2},\frac{3}{2},\frac{5}{2}$ (trajectories $a,b,c,d$ respectively).
On the left, $U\f=e^{i\frac {\pi}{2}\frac{\sigma_z}{2}}$ corresponds to a rotation about the $z$-axis.
On the right, $U\f=e^{i\frac {\pi}{4}\frac{\sigma_y}{2}}$ corresponds to a rotation about the $y$-axis.
}
\label{y-z traj det}
\end{figure}

Any target $U\f=(\psi\f,\theta\f,\phi\f)$ is reached by one and only one time-optimal control in $\Omega\d$.
In other words, the function $f\d$ is bijective on $\Omega\d$.
The time-optimal control generating $U\f$ is the unique control $u_{\Psi}$  solution of the equation
\begin{equation}
\label{optimal control}
u_{\Psi}=f\d^{-1}(\psi\f+n\cdot 4\pi)\cap{\Omega\d}
\end{equation}
for a certain $n\in\Z$.
Examples of the projected trajectories for two targets $U\f=e^{i\frac {\pi}{2}\frac{\sigma_z}{2}}$ and $U\f=e^{i\frac {\pi}{4}\frac{\sigma_y}{2}}$ under different detuning values $\Delta=0,\frac{1}{2},\frac{3}{2},\frac{5}{2}$ are depicted in Fig.~\ref{y-z traj det}.

\subsection{Control over $\SO(3)$}

Let us consider a target of the form $U=(\psi\f,\theta\f,\phi\f)$.
We are interested in answering the following question: \textit{Given $\Delta\neq0$, which target among $U$ and $-{U}$ is the fastest to generate?}
Or equivalently, if $\T(\Delta):= T(U,\Delta)-T(-{U},\Delta)$ is the ``difference time" function, \textit{for which values $\Delta$ does the $\T$ function change sign?}
Here, $T(U,\Delta)$ denotes the duration of the time optimal control generating $U$ under a detuning term $\Delta$.
In order to do so, we can of course proceed algorithmically by finding the time-optimal controls for $U$ and $-{U}$ respectively using the results of Section~\ref{sec5B} and then comparing the duration of the two corresponding controls.
But here, we aim at understanding under which conditions the function $\T$ changes sign at a specific value of $\Delta$.

There are two controls of particular interest denoted by $u_{\Psi_+}$ and $u_{\Psi_-}$, where $\Psi_{\pm}:=-\phi\f\pm\pi$.
They are special since they correspond to the only two controls ($\theta\f$, $\phi\f$ being fixed) which generate two opposite matrices $U:=(\Psi_+,\theta\f,\phi\f)$ and $-U:=(\Psi_-,\theta\f,\phi\f)$ in the same time $T(u_{\Psi_+})=T(u_{\Psi_-})$.
Let us split the full detuning domain $\R$ into two regions, $X$ and $\R/X$, where $X:=\{\Delta$ $|$ both $u_{\Psi_{+}},u_{\Psi_{-}}\in \Omega\d\}$.

\smallskip\noindent
When $\Delta\in X$, we can show that $u_{\Psi_+}$ is the optimal control generating $U\f$ if and only if  $u_{\Psi_-}$ is the optimal control generating $-U\f$ or vice-versa.
In particular, the values of $\Delta$ for which $u_{\Psi_+}$ and $u_{\Psi_-}$ are the optimal controls of $U\f$ and $-U\f$ are the ones for which the difference of time function $\T$ is equal to zero since $T(u_{\Psi_+})=T(u_{\Psi_-})$.
These values of $\Delta$ are easily found to be
$\Delta=-\frac{\phi\f+\psi\f\pm\pi+n\cdot 4\pi}{2T(u_{\Psi_{\pm}})}$ for a certain $n\in\Z$.

\smallskip\noindent
When $\Delta\notin X$, at least one of the two controls $u_{\Psi_{\pm}}$ is not time-optimal such that the difference of time function $\T$ is never zero on the set $\R / X$.
Since the time-optimal control domain $\Omega\d$ as well as the optimal control $u_{\Psi}$ of $U\f$ (or of $-U\f$) vary smoothly with $\Delta$ (this means that the angle $\Psi$ defining $u_{\Psi}$ is smooth), the function $\T$ also varies smoothly with respect to $\Delta$ except when the time-optimal control jumps from one bound to the other of $\Omega\d$.
This occurs when the optimal control for one of two unitary matrices $U$ or $-U$ jumps between $u_{\Psi_{min}}$ and $u_{\Psi_{max}}$ for an infinitesimal variation of $\Delta$.
We can summarize the previous discussion by the following Proposition.

\begin{proposition}
\label{general switching points}
For $\Delta\in X$, the function $\T$ changes sign if and only if $\Delta=-\frac{\phi\f+\psi\f\pm\pi+n\cdot 4\pi}{2T(u_{\Psi_{\pm}})}$ for a certain $n\in\Z$. Moreover, $\T=0$ at these points.
For $\Delta\in \R/X$, the function $\T$ changes sign if and only if the time optimal control for $U$ or $-U$ is $u_{\Psi_{min/max}}$.
\end{proposition}

\begin{figure}[ht]
\centering
\includegraphics[width=0.7\textwidth]{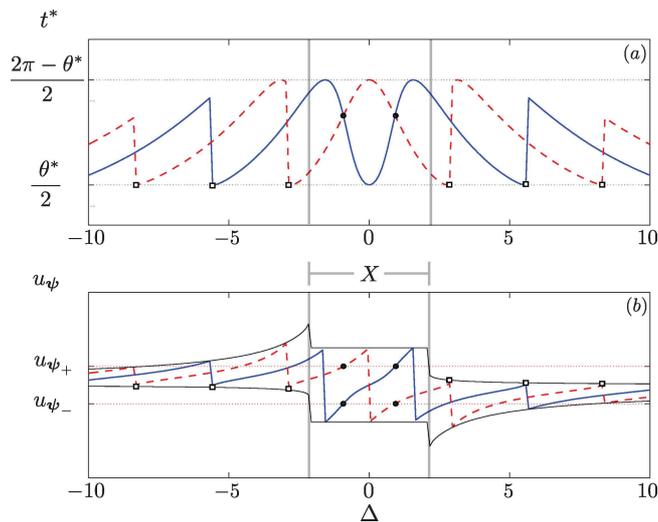}
\caption{(a) Time-optimal durations $T(U,\Delta)$ and $T(-U,\Delta)$ to generate $U=(0,2.2689,0)$ (blue/solid line) and $-U$ (red/dashed line) as a function of the detuning $\Delta$.
(b) Time-optimal controls $u_{\Psi}$ of $U$ (blue/solid line) and $-U$ (red/dashed line) respectively as a function of the detuning $\Delta$.
The region defined by the two black curves represents the time-optimal domain $\Omega\d$ for each value of $\Delta$.
The difference of time function $\T$ stops being continuous when both $u_{\Psi_+}$ and  $u_{\Psi_-}$ do not belong to $\Omega\d$  (domain $X$ defined by the two vertical lines).
The black dots and the white squares represent the values of $\Delta$ for which $\T$ changes sign for $\Delta\in X$ and $\Delta\in \R/X$ respectively.
}
\label{TvsDet_Iy}
\end{figure}

\begin{figure}
\centering
\includegraphics[width=0.7\textwidth]{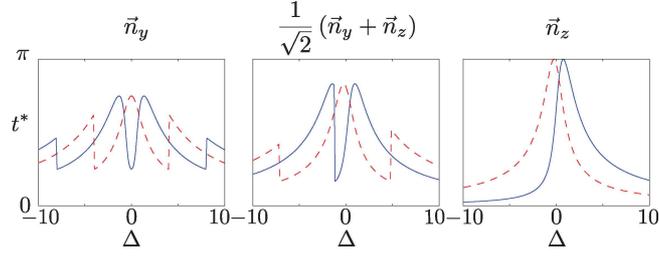}
\caption{Evolution of the minimum time $\tf$ as a function of $\Delta$ for rotations around the three axes
$\vec n_y, \frac{1}{\sqrt 2}(\vec n_y+\vec n_z)$ and $\vec n_z$.
The rotation angle is fixed to $\alpha=\pi/2$.}
\label{TvsrotDet}
\end{figure}

An example of application of Proposition~\ref{general switching points} is depicted in Fig.~\ref{TvsDet_Iy} for the unitary matrices $U=(\psi\f,\theta\f,\phi\f)=(0,1.9897,0)$ and $-U$.
The upper graph shows the optimal time for generating unitaries $U$ and $-U$ as a function of $\Delta$.
For each value of $\Delta$, the time-optimal domain $\Omega\d$ is given by the two enveloping black curves on the lower graph.
As long as both $u_{\Psi_{\pm}}\in\Omega_\Delta$ (region $X$), the optimal times $T(U,\Delta)$ and $T(-U,\Delta)$ vary continuously between the shortest and longest times and the two curves meet (black dots on Fig.\ref{TvsDet_Iy}~(a)) when their two time-optimal controls are exactly  $u_{\Psi_{\pm}}$ (black dots on Fig.\ref{TvsDet_Iy}~(b)).
As soon as one of $u_{\Psi_{\pm}}$ is not in $\Omega\d$ anymore ($\Delta\in\R/X$), the time functions stop crossing but jump one above the other when the optimal controls (curves of the lower graph) reach the  limits of $\Omega\d$.
Three other examples are given in Fig. \ref{TvsrotDet}.



\section{Conclusion}\label{sec6}
In this paper, we have investigated the time-optimal control of SU(2) quantum operations by a bounded external field with two components along the $x$- and $y$- directions. We have analyzed a control problem where the rotating wave approximation provides a valid simplification of the dynamics of the system. We have considered two different situations, with and without a constant detuning term. We have shown that geometric optimal control techniques provide a systematic way to attack such control problems, leading to the complete description of the pulse sequences. Furthemore, we have studied the basic model of a two-level quantum system in order to highlight the geometric structure of the control. Such results could be applied to more complicated systems such as the one presented in Ref. \cite{khaneja}, where the control of three coupled spins can be reduced to SU(2) operations.

\section{Appendix}
\label{app}

\subsection{Derivation of the dynamical Equations~\eqref{Dynamic theta_i det}}
\label{app0}
Let us write the quantum Hamiltonian in terms of the normalized variables $v_i:=\frac{\omega_i}{\omega_{max}}$ and $\Delta=\frac{\omega}{\omega_{max}}$ using the Quaternions notation:
$$\mathrm H
=v_x\sigma_x+v_y\sigma_y+\Delta\sigma_z
=-i(v_x\k+v_y\j+\Delta\i).
$$
Using the definition of $U={x_1}\id+{x_2}\i+{x_3}\j+{x_4}\k$ given in Eq.~\eqref{U quaternion dec}, the Schr\"odinger equation (see Eq.~\eqref{Dynamic U}) translates into

\begin{equation*}
\label{hamiltonian}
\begin{array}{c}
i\partial_t U(t)=\mathrm H(t)U(t) \\
\iff~\\
\dot x_1+\dot x_2\i+\dot x_3 \j+ \dot x_4 \k =\hspace{3.5cm}\\
\hspace{2cm} -(\Delta\i+v_y\j+v_x\k) \cdot ({x_1}\id+{x_2}\i+{x_3}\j+{x_4}\k)\\
\iff~\\
~~~\left(\begin{array}{c}
\dot {x_1}\\
\dot {x_2}\\
\dot {x_3}\\
\dot {x_4}
\end{array}\right)
=
L_{\mn iH}
\left(\begin{array}{c}
{x_1}\\
{x_2}\\
{x_3}\\
{x_4}\\
\end{array}\right)\\
\end{array}
\end{equation*}
%
where
$L_{iH}
=
\left(\begin{smallmatrix}
0   & \Delta & v_y & v_x\\
-\Delta & 0    & v_x & -v_y \\
-v_y & -v_x  & 0    & \Delta \\
-v_x & v_y & -\Delta  & 0 \\
\end{smallmatrix}\right).$
%
Now, noticing that the coordinates $x_i$ can be written in terms of the Hopf parameters as
\begin{equation*}
\begin{array}{rll}
x_1
&=\cos\theta_1\cos\theta_2,\\
x_2
&=\cos\theta_1\sin\theta_2,\\
x_3
&=\sin\theta_1\cos\theta_3,\\
x_4
&=\sin\theta_1\sin\theta_3,\\
\end{array}
\end{equation*}
where the variable $r=1$,
we deduce by straightforward calculations that

\begin{equation*}
\left(\begin{array}{c}
\dot x_1\\
\dot x_2\\
\dot x_3\\
\dot x_4
\end{array}\right)
=
T
\left(\begin{array}{c}
\dot {r}\\
\dot {\theta_1}\\
\dot {\theta_2}\\
\dot {\theta_3}\end{array}\right)
\end{equation*}
where $T=$
\begin{equation*}
\left(\begin{smallmatrix}
\cos\theta_1\cos\theta_2 & -\sin\theta_1\cos\theta_2& -\cos\theta_1\sin\theta_2 & 0\\
\cos\theta_1\sin\theta_2 & -\sin\theta_1\sin\theta_2& ~~\cos\theta_1\cos\theta_2 & 0\\
\sin\theta_1\cos\theta_3 &~~\cos\theta_1\cos\theta_3& 0 & -\sin\theta_1\sin\theta_3\\
\sin\theta_1\sin\theta_3 & ~~\cos\theta_1\sin\theta_3& 0 & ~~\sin\theta_1\cos\theta_3\\
\end{smallmatrix}\right).
\end{equation*}
The matrix $T$ is invertible when $\theta_1\neq \frac{n\pi}{2}$ for $n\in\mathbb N$ with inverse $T^{-1}=$
\begin{equation*}
\label{T-1}
\left(\begin{smallmatrix}
~~\cos\theta_1\cos\theta_2 & ~~\cos\theta_1\sin\theta_2 & ~~\sin\theta_1\cos\theta_3 & \sin\theta_1\sin\theta_3\\
-\sin\theta_1\cos\theta_2 &-\sin\theta_1\sin\theta_2 &  ~~\cos\theta_1\cos\theta_3 & \cos\theta_1\sin\theta_3\\
-\sec\theta_1\sin\theta_2 & ~~\sec\theta_1\cos\theta_2 & 0 & 0\\
0 & 0 & -\csc\theta_1\sin\theta_3 & ~~\csc\theta_1\cos\theta_3
\end{smallmatrix}\right).
\end{equation*}
The dynamics of the system can be described in terms of the Hopf variables:

\begin{equation*}
\left(\begin{array}{c}
\dot {r}\\
\dot {\theta_1}\\
\dot {\theta_2}\\
\dot {\theta_3}\end{array}\right)
= T^{-1}L_{iH}
\left(\begin{array}{c}
\cos \theta_1\cos \theta_2\\
\cos \theta_1\sin\theta_2 \\
\sin \theta_1\cos \theta_3\\
\sin \theta_1\sin \theta_3\\
\end{array}\right).
\end{equation*}
Knowing that $\dot r=0$, the above equality reduces to
\begin{equation*}
\left(\begin{array}{c}
\dot {\theta_1}\\
\dot {\theta_2}\\
\dot {\theta_3}\end{array}\right)
=
\left(
\begin{smallmatrix}
~~~-v_x\sin(\theta_2+\theta_3)-v_y\cos(\theta_2+\theta_3)\\
\tan\theta_1[~~v_x\cos(\theta_2+\theta_3)-v_y\sin(\theta_2+\theta_3)]-\Delta\\
\cot\theta_1[-v_x\cos(\theta_2+\theta_3)+v_y\sin(\theta_2+\theta_3)]-\Delta\\
\end{smallmatrix}
\right).
\end{equation*}

Using the definition of the control $v$ given in Eqs.~\eqref{real control} and the transformation given in Eqs.~\eqref{control change}, we obtain
\begin{equation*}
\begin{array}{ll}
u _1
    &=-v_0\sin(\mu+\theta_2+\theta_3)\\
&=-v_x\sin(\theta_2+\theta_3)-v_y\cos(\theta_2+\theta_3),\\
u_2
   &=-v_0\cos(\mu+\theta_2+\theta_3)\\
&=-v_x\cos(\theta_2+\theta_3)+v_y\sin(\theta_2+\theta_3),\\

\end{array}
\end{equation*}
and hence the dynamical equations~\eqref{Dynamic theta_i det}:
\begin{equation*}
\left(\begin{array}{c}
\dot {\theta_1}\\
\dot {\theta_2}\\
\dot {\theta_3}\end{array}\right)
=
\left(
\begin{array}{c}
~~u_1\\
-\tan\theta_1~u_2-{\Delta}\\
~~\cot\theta_1~u_2-{\Delta}\\
\end{array}
\right).
\end{equation*}

\subsection{Proof for the trajectory on the sphere}
\label{app1}
In order to prove this result, we will use the original definition of $\bar\phi:=\phi(0)-\beta(0)$ without replacing $\beta(0)$ by its value $-\frac\pi2$.
The reader can refer to Fig.~\ref{sphere parameters} for a visualization of the different variables encountered in the proof.
Let $\gamma:[0,t\f]\longrightarrow S^2$ be the path on the sphere described by the Euler angles $\theta(t)$ and $\phi(t)$ and let
$\vec\gamma(t):=(\sin\theta(t)\cos\phi(t),\sin\theta(t)\sin\phi(t),\cos\theta(t))$
denotes the vector in $\R^3$ defined by the point $\gamma(t)$.
We will first show that $\vec{\gamma}(t)\cdot\vec n\equiv\cos\bar\theta$ for all $t$ proving that $\gamma(t)$ lies in  a plan orthogonal to $\vec n$, and that the trajectory lies on a circle.

Neglecting some trivial algebra, let us first explicitely write
\begin{equation*}
\begin{array}{rl}
  \vec\gamma(t)\cdot \vec n
&= \sin\theta\sin\bar\theta(\cos\phi\cos\bar\phi+\sin\phi\sin\bar\phi)+\cos\theta\cos\bar\theta\\
&= \sin\theta\sin\bar\theta\cos(\phi-\bar\phi)+\cos\theta\cos\bar\theta\\
&= \sin\theta\sin\bar\theta\cos\beta+\cos\theta\cos\bar\theta\\
\end{array}
\end{equation*}
where we used the definition of $\bar\phi$ to write $\phi-\bar\phi=\phi-(\phi_0-\beta_0)=\beta$ where the last equality can be deduced from Eqs.~\eqref{cst motion}.
Factorizing by $\cos\bar\theta$ and using the definition of $\bar\theta$, the preceding equality becomes
\begin{equation*}
\begin{array}{rl}
  \vec\gamma(t)\cdot \vec n
&= \cos\bar\theta(\tan\bar\theta\sin\theta\cos\beta+\cos\theta)\\
&= \cos\bar\theta(\frac{1}{\p_2}\sin\theta\cos\beta+\cos\theta).\\
\end{array}
\end{equation*}

Now, as $\beta$ is the angular parameter for $-u$ where $u$ is the rotated control (see Eq.~\eqref{control change}), we write $\cos\beta=-u_2=\p_2\tan\theta_1$ using the definition of the controls given in Eqs.~\eqref{controls}.
As $\theta=2\theta_1$, we finally get:
\begin{equation*}
\begin{array}{rl}
  \vec\gamma(t)\cdot \vec n
&= \cos\bar\theta(\frac{1}{\p_2}\sin(2\theta_1)\cdot\p_2\tan\theta_1+\cos(2\theta_1))\\
                             &= \cos\bar\theta(2\sin\theta_1\cos\theta_1\tan\theta_1+(\cos^2\theta_1-\sin^2\theta_1))\\
                             &= \cos\bar\theta(2\sin^2\theta_1+\cos^2\theta_1-\sin^2\theta_1)\\
                             &= \cos\bar\theta(\sin^2\theta_1+\cos^2\theta_1)\\

&= \cos\bar\theta.\\
\end{array}
\end{equation*}

It remains to show that the trajectory runs with constant velocity $\|\vec{\dot\gamma}(t)\|=2$ on the circle.
We will use two relations, namely $\dot\theta=2\dot\theta_1=2u_1$ as well as $\dot\phi=-u_2(\tan\theta_1+\cot\theta_1)$ obtained from Eq.~\eqref{Dynamic theta_i det} and the definition of $\dot\phi=\dot\theta_2-\dot\theta_3$.
Having already simplified the expression for $\| \vec{\dot\gamma}(t)\|^2$, we obtain the following sequence of equalities leading to the desired relation:
\begin{equation*}
\begin{array}{rcl}
\| \vec{\dot\gamma}(t)\|^2
&=&[\dot\theta\cos\theta]^2+[\dot\phi\sin\theta]^2+[\dot\theta\sin\theta]^2\\
&=&[\dot\theta]^2+[\dot\phi\sin\theta]^2\\
&=&[\dot\theta]^2+[\dot\phi\sin\theta]^2\\
&=&[2u_1]^2+[-u_2(\tan\theta_1+\cot\theta_1)\sin2\theta_1]^2\\
&=&[2u_1]^2+[-u_2(\tan\theta_1+\cot\theta_1)2\sin\theta_1\cos\theta_1]^2\\
&=&[2u_1]^2+[-2u_2]^2\\
&=&4\|u\|^2\\
&=&4,\\
\end{array}
\end{equation*}
which completes the proof.

\subsection{Idea of the proof of Proposition~\ref {optimal domain}}
\label{app2}

\begin{figure}
\centering
\includegraphics[width=0.45\textwidth]{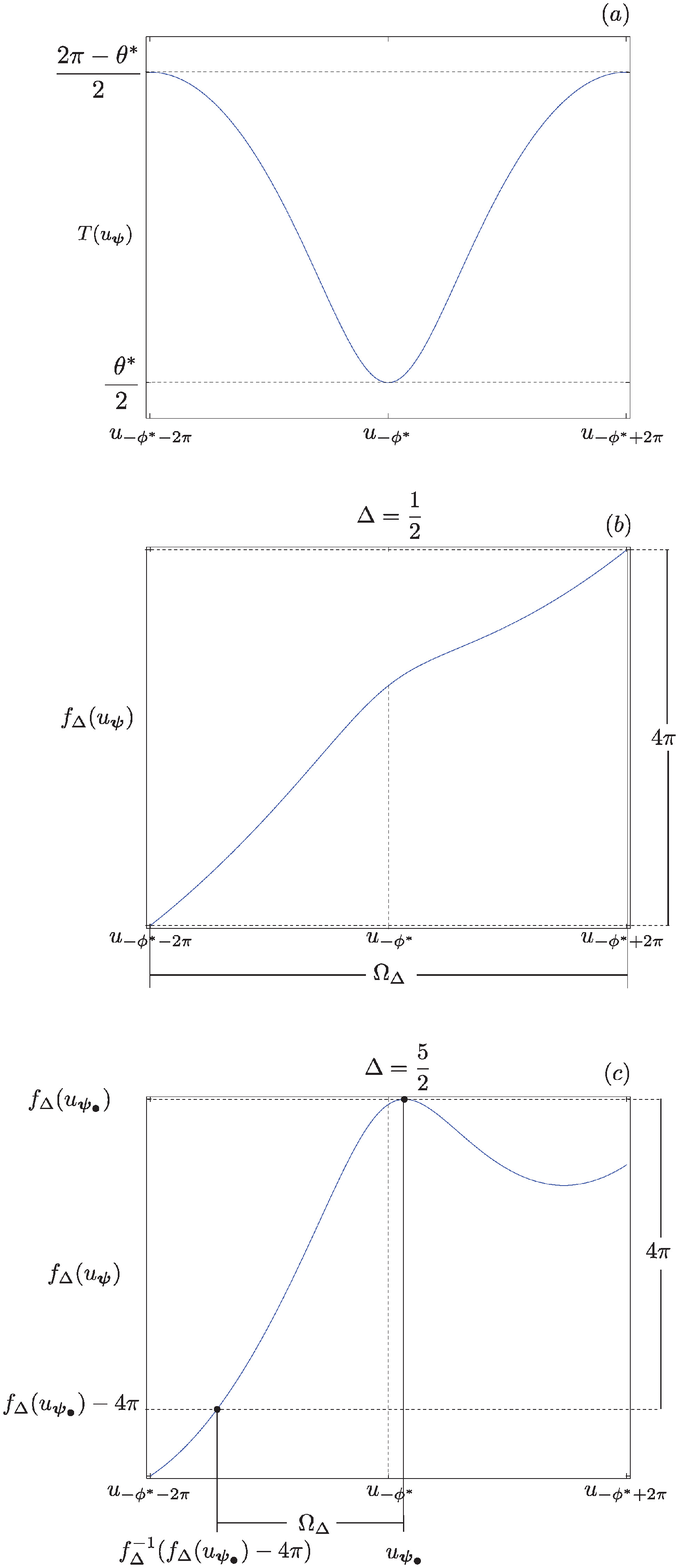}
\caption{(a) Plot of the time duration $T(u_{\Psi})$ of the controls $u_{\Psi}$ for $\Psi\in[-\phi\f-2\pi,-\phi\f+2\pi]$.
(b,c) Plot of the restriction $f\d$ of the end-point mapping for detuning values
$\Delta=\frac 12$ and $\frac32$ respectively.
The panel (b) illustrates the case $\Delta\leq|\tan(\frac{\theta\f}{2})|$ where the optimal domain $\Omega\d$ consists of the full domain $[u_{-\phi\f-2\pi},u_{-\phi\f+2\pi}]$.
The panel (c) depicts the case $\Delta>|\tan(\frac{\theta\f}{2})|$ where the optimal domain $\Omega\d\subset[u_{-\phi\f-2\pi},u_{-\phi\f+2\pi}]$. In this example, the angular parameters $\theta\f=2.2689$ and $\phi\f=0$ are fixed and taken as in Fig.~\ref{TvsDet_Iy}.}
\label{TvsPSI}
\end{figure}

\noindent
In this section, we give the main lines of the proof of Proposition~\ref {optimal domain}, which will be proved rigorously in an upcoming mathematical publication \cite{newmath}.
Recall that $\phi\f$ and $\theta\f$ are fixed such that the regular controls which are considered are uniquely denoted $u_{\Psi}$.
Let us first start with some claims about the control duration function
$T:u_{\Psi} \mapsto T(u_{\Psi})$
that we are not going to prove in the present paper.
\begin{enumerate}[\rm (i)]
\item $T(u_{\Psi})$ is continuous in $\Psi$.
\item $T(u_{\Psi})$ is concave, symmetric in its minimum $\Psi=-\phi\f$, and reaches its maxima at $\Psi=-\phi\f\pm 2\pi$.
\item $\frac{d T(u_{\Psi})}{d\Psi}=\frac{\p_2(u_{\Psi})}{2}$
\end{enumerate}
where $\p_2(u_{\Psi})$ is the value of the adjoint variable $\p_2$ along the trajectory corresponding to the optimal control $u_{\Psi}$.
The shape of the time function described in claim (i) is illustrated in Fig.~\ref{TvsPSI}~(a).

We define the time-optimal domain
$$\Omega\d=[u_{\Psi_{min}},u_{\Psi_{max}}]=\{u_{\Psi}|\Psi\in[\Psi_{min},\Psi_{max}]\}$$ corresponding to the set of all the time-optimal controls in the presence of a detuning term $\Delta$,  where $\theta\f$ and $\phi\f$ are fixed.

First note that since $T(u_{-\phi\f+2\pi})=T(u_{-\phi\f-2\pi})$ (claim (ii)), $f\d(u_{-\phi\f+2\pi})-f\d(u_{-\phi\f-2\pi})=4\pi$ such that the function $f\d$ is at least surjective on the full domain
$[u_{-\phi\f+2\pi},u_{-\phi\f+2\pi}]$ (see Fig.~\ref{TvsPSI}~(b) and (c)).
To study the injectivity of $f\d$, consider its derivative:

$$
\frac{df\d(u_{\Psi})}{d\Psi}
=1-2\Delta\frac{d T(u_{\Psi})}{d\Psi}
=1-\Delta\p_2
$$
where we used the claim (iii) to deduce the last equality.
From Proposition~\ref{trajectory circle}, we know that $|\p_2|\leq\left|\cot\left(\frac{\theta\f}{2}\right)\right|$ such that for $|\Delta|\leq|\tan(\frac{\theta\f}{2})|$, the derivative $\frac{df\d(u_{\Psi})}{d\Psi}$ never vanishes or does on the boundary of its domain.
In this case, the function $f\d$ is then injective (and so bijective) on the full domain $[u_{-\phi\f-2\pi},u_{-\phi\f+2\pi}]$ which then consists of the searched time-optimal domain $\Omega_{\Delta}$.
This case is depicted in Fig.~\ref{TvsPSI}~(b).\\
\indent For $|\Delta|>|\tan(\frac{\theta\f}{2})|$, let $\psi\bl$ denote the point (among possibly two) which is the closest to $-\phi\f$ and for which $\left.\frac{df\d(u_{\Psi})}{d\Psi}\right|_{\Psi\bl}=0$.
Suppose without loss of generality that $\Delta>0$.
One can show that $\Psi\bl\geq-\phi\f$ and that $f\d$ strictly increases on an "interval" $[u_{\Psi}, u_{\Psi\bl}]$ for which $f\d(u_{\Psi\bl})-f\d(u_{\Psi})\geq 4\pi$.
One then selects the smallest interval $[u_{\Psi_{min}},u_{\Psi\bl}]$ on which $f\d$ is bijective.
This unique interval having $u_{\Psi\bl}$ as an upper bound is the interval the most centered in $-\phi\f$ on which $f\d$ is bijective.
It corresponds to the controls having shortest durations due to the claim (ii).
One finally directly computes $\Omega\d=[f\d^{-1}(f\d(u_{\Psi\bl})-4\pi),u_{\Psi\bl}]$.
This case is depicted in Fig.~\ref{TvsPSI}~(c).

\end{document}